\documentclass[aps,prb,superscriptaddress,letterpaper,amsmath,amssymb,twocolumn,10pt,floatfix]{revtex4-2}
\usepackage{graphicx}
\usepackage[ansinew]{inputenc} 
\usepackage{array} 
\usepackage{xcolor} 
\usepackage{amsxtra}
\usepackage{amstext}
\usepackage{amsthm} 
\usepackage{latexsym}
\usepackage{verbatim} 
\usepackage{bm} 
\usepackage{hyperref} 

\usepackage{color}
\usepackage{amssymb}

\usepackage{graphicx}
\usepackage[ansinew]{inputenc} 
\usepackage{array} 
\usepackage{amsxtra}
\usepackage{amstext}
\usepackage{amsthm} 
\usepackage{latexsym}
\usepackage{IEEEtrantools} 
\usepackage{verbatim} 
\usepackage{bm} 
\usepackage{hyperref} 

\begin{document}


\title{ Skyrmion topology of scalar waves in real and momentum space }

\author{Nai H. Kwong}
\affiliation{Wyant College of Optical Sciences, University of Arizona, Tucson, AZ 85721}

\author{Ewan M. Wright}
\affiliation{Wyant College of Optical Sciences, University of Arizona, Tucson, AZ 85721}

\author{Jan Wingenbach}
\affiliation{Department of Physics and Center for Optoelectronics and Photonics Paderborn (CeOPP), Paderborn University, 33098 Paderborn, Germany}
\affiliation{Institute for Photonic Quantum Systems (PhoQS),
Paderborn University, 33098 Paderborn, Germany}

\author{Roman Lebs}
\affiliation{Department of Physics and Center for Optoelectronics and Photonics Paderborn (CeOPP), Paderborn University, 33098 Paderborn, Germany}
\affiliation{Institute for Photonic Quantum Systems (PhoQS),
Paderborn University, 33098 Paderborn, Germany}

\author{Harald Giessen}
\affiliation{Physikalisches Institut, Research Center SCoPE, and Integrated Quantum Science and Technology Center (IQST), University of Stuttgart, 70569 Stuttgart, Germany}
\affiliation{Wyant College of Optical Sciences, University of Arizona, Tucson, AZ 85721}


\author{Stefan Schumacher}
\affiliation{Wyant College of Optical Sciences, University of Arizona, Tucson, AZ 85721}
\affiliation{Department of Physics and Center for Optoelectronics and Photonics Paderborn (CeOPP), Paderborn University, 33098 Paderborn, Germany}
\affiliation{Institute for Photonic Quantum Systems (PhoQS),
Paderborn University, 33098 Paderborn, Germany}

\author{Rolf Binder}
\affiliation{Wyant College of Optical Sciences, University of Arizona, Tucson, AZ 85721}
\affiliation{Department of Physics, University of Arizona, Tucson, AZ 85721}


\date{\today}

\begin{abstract}
Generalizations of magnetic skyrmions to the context of other areas of physics, such as plasmonics, optics (including microcavity polaritons) and even water waves have become an important research topic. 
However, the generalization of the skyrmion number in such systems is still the subject of active research. While magnetic skyrmions textures are inherently formed by vectors (magnetic spin vectors), recent developments in other areas of physics have generalized the concept of skyrmions and skyrmionic patterns to the case in which scalar fields are used to define 3-component pseudo-spin vectors.  This gives rise to a wide variety of skyrmion-like textures, depending on the detailed functional forms of the scalar fields.
For example, steady-state outgoing waves in driven-dissipative polariton systems fail to have a well-defined global skyrmion number in real (configuration) space but can have one in momentum space.
We classify the skyrmion integral for various scalar fields with different asymptotic behaviors in real space as well as momentum space, and in both cases give conditions for the skyrmion number not to exist, or be integer, half-integer or non-integer.
Examples of physical systems to which our analysis can be applied include polariton condensates, water waves and gain-guided lasers.
\end{abstract}

\maketitle


\section{Introduction}

Spatial textures of vector fields   play a central role across condensed matter, photonics, and fluid dynamics. In magnetic and spin systems, the concept of magnetic skyrmions is well established (e.g. 
\cite{roessler-etal.2006,nagaosa-etal.2013,cook.2023,tokura2020magnetic,drissi.2022arxiv}). These spin textures are called double-twisted, since the spins rotate simultaneously in the radial and azimuthal direction, and the textures that look like concentric rings are called target skyrmions 
\cite{leonov-etal.2014}.
The physics of magnetic skyrmion formation is still a topic of current interest \cite{liefferink-etal.2026}.
Other areas where the concept of magnetic skyrmions has been adopted and extended include
optics~\cite{Du2019,doi:10.1126/science.aau0227,Shen2024,lpor.202501427,Krol:21,peters-etal.2026},  Bose-Einstein condensates (BECs)~\cite{ohmi-machida.1998,PhysRevLett.81.742}, phonons~\cite{cao2023observation},
topological encoding \cite{zhang-etal.2026skyrmions}, spin-Hall effect in semiconductors \cite{flayac-etal.13prl}, and spinor exciton-polariton condensates~\cite{CHENG2024129600}.
Further examples of skyrmion textures in other areas of physics include
water waves
\cite{smirnova2024water},  plasmonic nanostructures~\cite{davis2020ultrafast,schwab2024plasmonic,schwab2025skyrmion}, hydrodynamic surface waves~\cite{wang2025topological,che2026twisted},  surface phonon polariton systems~\cite{schwab2026tunable,mangold2026phonon},
and exciton-polaritons in semiconductor microcavities \cite{wingenbach-etal.2026skyrmions}.

The various extensions of the concept of skyrmions have made it necessary to revisit the topological properties. The  skyrmion number for magnetic skyrmions is usually an integer reflecting a global topological invariant or topological charge, which has analogies to the original skyrmion concept from field theory \cite{skyrm.1961}.
However,  magnetic skyrmions can also exhibit non-integer skyrmion numbers, see for example Ref. \cite{leonov-etal.2014}.

For our analysis it is important that the recent extensions of the concept of magnetic skyrmions also do not generally involve integer-valued skyrmion numbers and therefore global topological invariants. In ideal isolated  magnetic skyrmions without spatial confinement, the skyrmion number has been found to be equivalent to the Chern number in the case of one-electron Hamiltonians represented by $2 \times 2$ matrices (this equivalence breaks down for larger Hamiltonian dimensions \cite{cook.2023}).
Recent work has emphasized that the conventional integer-valued skyrmion numbers require appropriate boundary conditions that allow compactification of the physical domain,
and has introduced generalized topological invariants in momentum space \cite{neuhaus-etal-1.2026arxiv,neuhaus-etal-2.2026arxiv}
and 
for  optical fields propagating  in three dimensions 
\cite{wang-etal.2026,zhang-etal.2026skyrmions}, respectively.

In the case of scalar-field based skyrmions, such as water waves and polariton condensates, it has been pointed out in, for example, Refs.
\cite{neuhaus-etal-1.2026arxiv,neuhaus-etal-2.2026arxiv,wingenbach-etal.2026skyrmions}, that complications arise from the fact that in scalar wave systems the spin texture over the domain (for example the real configuration space of the wave) is infinite. 
Other examples of scalar fields that can support skyrmion-like textures include
the familiar Gaussian beams and laser modes of gain guided lasers, for example 
\cite{petermann.1979,siegman.1989,salin-etal.1992} (the concept of gain-guided lasers has also been extended to polariton system in, for example, Refs.
\cite{wouters-carusotto.2007,alyatkin-etal.2021}).
Their amplitude decay and radial phase determine whether the associated pseudo-spin texture covers a hemisphere, forms skyrmion cells, or admits a global skyrmion number.

While in `simple' magnetic skyrmions  the spin state is uniformly (in all directions) the same when the distance from its center becomes large (or infinite),
this need not be true in more general skyrmion-like textures. For example, the spin (more correctly the pseudo-spin) may keep oscillating in the limit of infinite radial distance from the skyrmion's center. 
Mathematically, this  means, as alluded to above,  that the domain cannot be compactified (in the sense that a uniform limit of the skyrmion function at infinity may not exist), and the notion of a degree of a map (from the domain to the sphere $S^2$ where the skyrmion pattern lives) may not be applicable. 

Since, in principle, there can be a large variety of scalar fields taken to be candidates for the formation of skyrmion-like textures, for example constructed along the lines of Refs. \cite{neuhaus-etal-1.2026arxiv,neuhaus-etal-2.2026arxiv,wingenbach-etal.2026skyrmions},  the question arises how a skyrmion number (or generalized skyrmion number) is related to the functional form of a given candidate scalar field. In other words, given a 2-dimensional domain (either real or momentum space), and given a real-valued  and normalized pseudo-spin function defined over that domain,  what is the expected skyrmion number? This is the question we want to address in the following.  In this paper, we restrict ourselves to the case of scalar functions that depend only on the radial coordinate, not on the azimuth angle (in other words, functions corresponding to zero angular momentum).

One main result is that, for the scalar-field construction considered here, the skyrmion integral is determined by the asymptotic behavior of the scalar field and its radial derivative. This allows us to classify when a global skyrmion number exists and when it is integer, non-integer, or undefined. We further find a pronounced difference between real and momentum space; scalar waves with infinitely continued radial oscillations may fail to possess a global real-space skyrmion number, while their Fourier-transformed version can support a well-defined integer momentum-space topology.

In the main part of the paper,  Sec. \ref{Sec:analysis-skyrmion-number},   we analyze scalar-field based generalized skyrmion numbers for skyrmion textures in configuration space, Sec.  
\ref{Sec:Scalar-field-based-real-space-skyrmions}, and momentum space, Sec.
\ref{Sec:Scalar-field-based-momentum-space-skyrmions}. and we provide conclusions in Sec. \ref{Sec:conclusion}

\begin{figure}[t]
        \includegraphics[width=0.5 \textwidth]{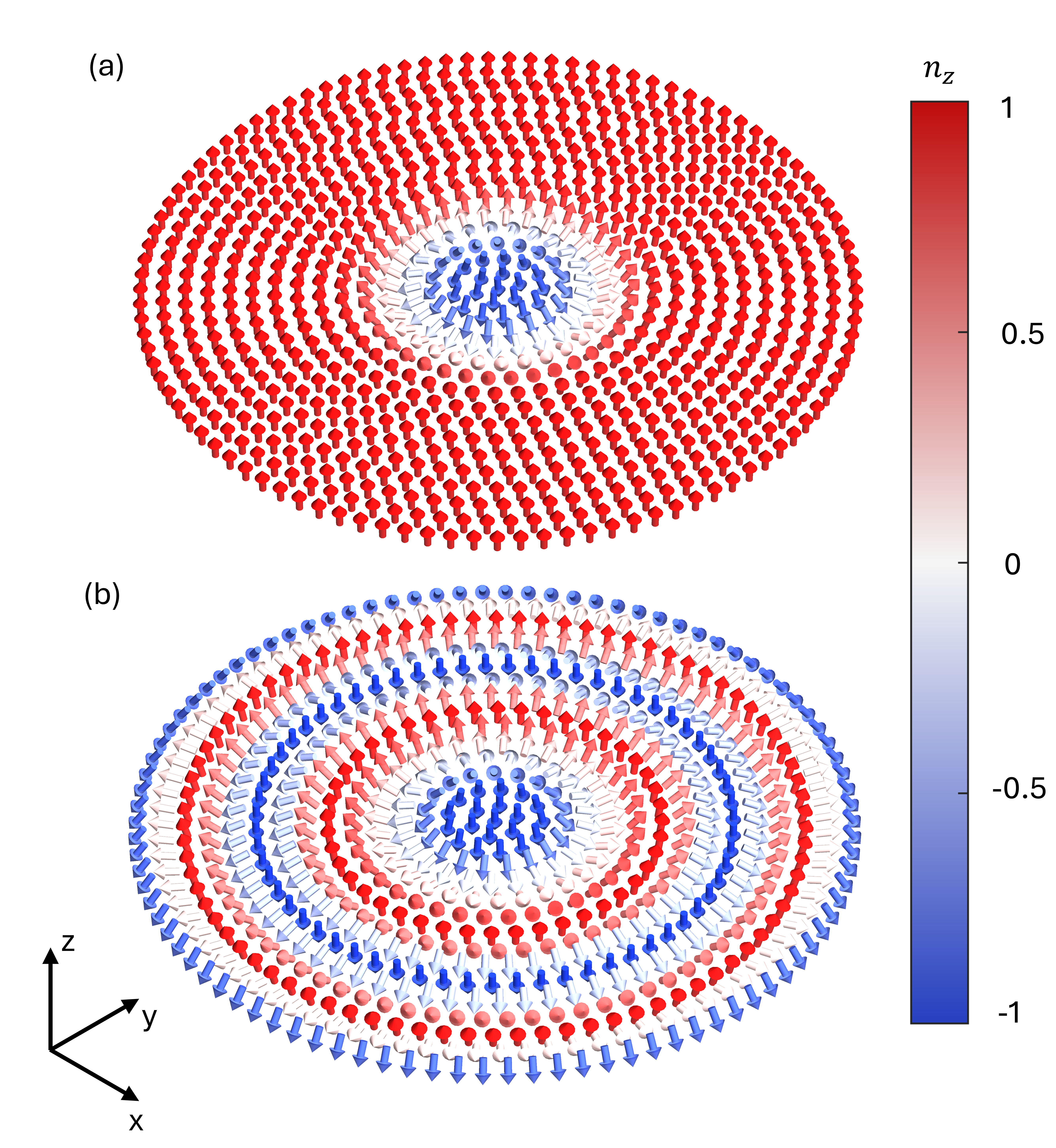} 
        \begin{center}
            \caption{Comparison of a conventional magnetic skyrmion and a scalar-wave skyrmion-like texture, according to Eq. \eqref{I.3} with $\Phi(\theta)=  m \theta +\gamma$ (here the vorticity $m=1$ and helicity $\gamma = 0$).   Colors represent the out-of-plane component $n_z$ of the normalized (pseudo-)spin field. a) In a conventional magnetic skyrmion, the spin field approaches a uniform orientation far from the core, allowing the two-dimensional domain to be compactified. b) In scalar-wave systems, the pseudospin may continue to oscillate with radial distance. The absence of a uniform asymptotic state prevents compactification of the domain.}
            \label{fig:schematic-skyrmion-texture}
        \end{center}
\end{figure}

\section{Analysis of  skyrmion integrals and skyrmion numbers.}

\label{Sec:analysis-skyrmion-number}

In this section, we discuss skyrmion numbers of single, isolated skyrmions constructed from circularly symmetric, real scalar functions. We consider 3D vector fields defined in a 2D domain. The domain can be configuration space  ($r$-space) or momentum space ($k$-space).

\subsection{Skyrmions in configuration space}

\label{Sec:skyrmion-real-space}

\subsubsection{Skyrmion number}
For clarity, let us begin by stating the well-known definition of the skyrmion number for an isolated skyrmion. For the two-dimensional (2D) domain we use the entire plane in $r$-space $\mathbb{R}^{2}$ 
(with 2D vectors $\mathbf{r}=(x, y)=(r \cos \theta, r \sin \theta))$, which for convenience we consider as disks of radius $r$ in the limit $\mathrm{r} \rightarrow \infty$. Then, for a given 3D real vector field $\mathbf{n}(x, y)$ of unit magnitude in this domain, the skyrmion number is given by the integrals
\begin{eqnarray}
\tilde{Q}(r)=\frac{1}{4 \pi} \int_{0}^{r} d r^{\prime} r^{\prime} \int_{0}^{2 \pi} d \theta \mathbf{n} \cdot\left[\frac{\partial \mathbf{n}}{\partial x} \times \frac{\partial \mathbf{n}}{\partial y}\right] \label{I.1.a} \\
=\frac{1}{4 \pi} \int_{0}^{r} d r^{\prime} \int_{0}^{2 \pi} d \theta \mathbf{n} \cdot\left[\frac{\partial \mathbf{n}}{\partial r^{\prime}} \times \frac{\partial \mathbf{n}}{\partial \theta}\right] \label{I.1}
\end{eqnarray}
and
\begin{eqnarray}
Q=\lim_{r \rightarrow \infty} \tilde{Q}(r)=\frac{1}{4 \pi} \int_{0}^{\infty} d r^{\prime} \int_{0}^{2 \pi} d \theta \mathbf{n} \cdot\left[\frac{\partial \mathbf{n}}{\partial r^{\prime}} \times \frac{\partial \mathbf{n}}{\partial \theta}\right] \label{I.2}
\end{eqnarray}
where under the integral it is implied that $\mathbf{n}=\mathbf{n}\left(r^{\prime}, \theta\right)$. The algebra validating  Eq. \ref{I.1} is shown in Appendix \ref{Sec:appendix-nagaosa-formula}. In the following, we will call $\tilde{Q}(r)$ the skyrmion integral and $Q$ the global skyrmion number. Since our analysis aims at understanding similarities and differences of generalized skyrmion textures with magnetic skyrmions, we will use the term pseudo-spin for the normalized 3-component vector field $\mathbf{n}(x,y)$   ($\mathbf{n}(k_x,k_y)$) in configuration (momentum) space.

\subsection{`Hedgehog' vector fields}
We consider a class of unit vector fields which are separable in their dependence on $r$ and $\theta$ according to
\begin{equation}
\mathbf{n}(r, \theta)=\left[\begin{array}{c}
\cos \Phi(\theta) \sin R(r)     \label{I.3} \\
\sin \Phi(\theta) \sin R(r) \\
\cos R(r)
\end{array}\right]
\end{equation}
where $R(r)$ and $\Phi(\theta)$ are any real-valued (smooth) functions. For this vector field class, the integration in Eqs. \eqref{I.1}  and \eqref{I.2}  can be done analytically (algebra shown in Appendix \ref{Sec:appendix-nagaosa-formula}) giving
\begin{eqnarray}
& \tilde{Q}(r)=-\frac{1}{4 \pi}\left[\cos R\left(r^{\prime}\right)\right]_{r^{\prime}=0}^{r^{\prime}=r}[\Phi(\theta)]_{\theta=0}^{\theta=2 \pi}  \label{I.4a}\\
& Q=\lim_{r \rightarrow \infty} \tilde{Q}(r) \label{I.4b}
\end{eqnarray}

The class of Eq. \eqref{I.3} is frequently used in the literature, and the result in Eq. \eqref{I.4a}  is known (see, for example Ref. \cite{nagaosa-etal.2013}, p. 901; the factor $-\frac{1}{4 \pi}$ seems to be missing in its expression for $Q$; see also \cite{peters-etal.2026}). The vector fields of interest to us here belong to this class.

We specialize to a subset of Eq. \eqref{I.3}
\begin{equation}
\mathbf{n}(r, \theta)=\frac{1}{\mathcal{N}(r)}\left[g_{2}(r) \hat{r}+g_{1}(r) \hat{z}\right] \label{I.5a}
\end{equation}
\[
=\frac{1}{\mathcal{N}(r)}\left[\begin{array}{c}
g_2(r)\cos \theta  \label{I.5b}\\
g_2(r)\sin \theta \\
g_{1}(r)
\end{array}\right]
\]
where $g_{1}(r)$ and $g_{2}(r)$ are real, circularly symmetric, scalar functions and the normalization factor is
\begin{equation}
\mathcal{N}(r)=\sqrt{\left[g_{1}(r)\right]^{2}+\left[g_{2}(r)\right]^{2}} \label{I.6}
\end{equation}

In the notation of Eq. \eqref{I.3}, the functions are
\begin{equation}
\Phi(\theta)=\theta, \quad \cos R(r)=\frac{g_{1}(r)}{\mathcal{N}(r)}, \quad \sin R(r)=\frac{g_{2}(r)}{\mathcal{N}(r)} \label{I.7}
\end{equation}

Substituting into Eq. \eqref{I.5a} gives the skyrmion integral and the global skyrmion number as
\begin{equation}
\tilde{Q}(r)=\frac{1}{2}\left[n_{z}(0)-n_{z}(r)\right], \quad Q=\frac{1}{2}\left[n_{z}(0)-\lim_{r \rightarrow \infty} n_{z}(r)\right] \label{I.8}
\end{equation}
with
\begin{equation}
n_{z}(r)=\frac{g_{1}(r)}{\sqrt{\left[g_{1}(r)\right]^{2}+\left[g_{2}(r)\right]^{2}}}=\frac{\operatorname{sign}\left[g_{1}(r)\right]}{\sqrt{1+\left[\frac{g_{2}(r)}{g_{1}(r)}\right]^{2}}} \label{I.9}
\end{equation}

The global skyrmion number $Q$ is well-defined if $\lim_{r \rightarrow \infty} n_{z}(r)$ exists, but it is not necessarily integer valued.

\subsection{Scalar-field based real-space skyrmions}
\label{Sec:Scalar-field-based-real-space-skyrmions}

In the case of skyrmion fields derived from scalar fields, which is used, for example, in polaritonic skyrmions 
\cite{wingenbach-etal.2026skyrmions} 
, one uses a further specialization of the form of the scalar functions
\begin{equation}
g_{1}(r)=f(r), \quad g_{2}(r)=\frac{\partial f}{\partial r} \label{I.10}
\end{equation}
where $f(r)$ is the real part of the coherent complex polariton wavefunction $\psi(r)$. Explicitly,
\begin{equation}
\mathbf{n}(r, \theta)=\frac{1}{\sqrt{f^{2}+\left(\frac{\partial f}{\partial r}\right)^{2}}}\left[\begin{array}{c}
\cos \theta \frac{\partial f}{\partial r}  \label{I.11}\\
\sin \theta \frac{\partial f}{\partial r}     \\
f
\end{array}\right] \quad, \quad f(r)=\operatorname{Re} \psi(r)
\end{equation}
which leads to
\begin{eqnarray}
\tilde{Q}(r)=\frac{1}{2}\left[n_{z}(0)-n_{z}(r)\right] \label{I.12a}\\
 \quad Q=\frac{1}{2}\left[n_{z}(0)-\lim _{r \rightarrow \infty} n_{z}(r)\right]  \label{I.12}\\
n_{z}(r)=\frac{f}{\sqrt{f^{2}+\left(\frac{\partial f}{\partial r}\right)^{2}}} \label{I.13a} \\
=\frac{f}{|f|} \frac{1}{\sqrt{1+\left(\frac{1}{f} \frac{\partial f}{\partial r}\right)^{2}}} \label{I.13}
\end{eqnarray}

Examples of 3-component skyrmion vector fields in other physical systems, where the in-plane components are given by the gradient of the third vector component, include water waves, 
see, for example, Eq. (9) of Ref. \cite{smirnova2024water}) and also Ref. \cite{wang2025topological}, and plasmonics skyrmion waves, see, for example, Eq. (1) of Ref. \cite{neuhaus-etal-2.2026arxiv}.

\subsubsection{ Asymptotic behavior of  the skyrmion number}

The previous subsections show that, for a hedgehog field configuration, the large $r$ behavior of the $z$-component of the unit vector field determines whether the global skyrmion number $Q$ is well-defined or not. In this subsection, we analyze the effects of this asymptotic behavior of $n_{z}(r)$ for vector fields encountered in the case of scalar-field based skyrmions.

Eq. \eqref{I.13} shows that the analysis of $n_{z}(r)$ is reduced to that on $\operatorname{sign}(f)$ and $\frac{1}{f} \frac{\partial f}{\partial r}$. We consider various types of the function $f(r)$ for this purpose. We assume that (i) $f(r)$ is a smooth function that decays to zero at infinity, and (ii) $f(0)$ is finite while $\left.\frac{\partial f}{\partial r}\right|_{r=0}=0$. The zero-derivative condition of assumption (ii) is satisfied by functions $f(r)$ that represent smooth, circularly symmetric functions in 2D. Assumption (ii) on $f(r)$ implies that $n_{z}(0)$ is given by the sign of $f(0)$:
\begin{equation}
n_{z}(0)=\frac{f(0)}{|f(0)|}= \pm 1 \label{I.14}
\end{equation}

We divide the functions $f(r)$ into two groups according to whether the function crosses zero a finite number or an infinite number of times over $r \in(0, \infty)$:

\underline{\textbf{(a)} $f(r)$ crosses zero a finite number of times.}

Denote the number of zero-crossings by $n_{c}$. Assumption (i) restricts the functions we consider to those decaying to zero. The asymptotic behavior of $\frac{1}{f} \frac{\partial f}{\partial r}$ then depends on the relative decay rate of $f(r)$ and $\frac{\partial f}{\partial r}$ beyond the last zero crossing. We divide the cases as follows: (In the following, when $\pm$ or $\mp$ appears, the upper (lower) sign corresponds to $f(0)$ being positive (negative).)

\textbf{(a.1)} $\frac{1}{f} \frac{\partial f}{\partial r} \rightarrow 0$ when $r \rightarrow \infty$.

In this case, we have
\begin{eqnarray}
\lim _{r \rightarrow \infty} n_{z}(r)=\lim _{r \rightarrow \infty} \frac{f(r)}{|f(r)|} & = \pm(-1)^{n_{c}}  \label{I.15}\\
Q= \pm \frac{1}{2}\left[1-(-1)^{n_{c}}\right] & = \begin{cases}0 & \text { for even } n_{c} \\
\pm 1 & \text { for odd } n_{c}\end{cases} \label{I.16}
\end{eqnarray}

An example of this group is a power-law decaying function
\begin{equation}
f(r) \sim \frac{1}{r^{n}}, n>0 \Rightarrow \frac{1}{f} \frac{\partial f}{\partial r} \sim-\frac{n}{r} \rightarrow 0 \text { as } r \rightarrow \infty \label{I.17}
\end{equation}

\textbf{(a.2)} $\left|\frac{1}{f} \frac{\partial f}{\partial r}\right| \rightarrow \infty$ when $r \rightarrow \infty$.
In this case, $\lim _{r \rightarrow \infty} n_{z}(r)=0$, and
\begin{equation}
Q= \pm \frac{1}{2} \label{I.18}
\end{equation}

An example of this group is a Gaussian
\begin{equation}
f(r) \sim e^{-a r^{2}}, a>0 \Rightarrow \frac{1}{f} \frac{\partial f}{\partial r} \sim-2 a r \rightarrow \infty \text { as } r \rightarrow \infty \label{I.19}
\end{equation}

In terms of Gaussian beams, this corresponds to the position of the beam's waist.

\textbf{(a.3)} $\frac{1}{f} \frac{\partial f}{\partial r} \rightarrow b$ when $r \rightarrow \infty$ where $b$ is a non-zero constant.

In this case,
\begin{equation}
\lim _{r \rightarrow \infty} n_{z}(r)= \pm(-1)^{n_{c}} \frac{1}{\sqrt{1+b^{2}}} \label{I.20}
\end{equation}
and
%
%
\begin{align}
Q & = \pm \frac{1}{2}\left[1-(-1)^{n_{c}} \frac{1}{\sqrt{1+b^{2}}}\right] \\
& \in \begin{cases}(-1 / 2,0) \cup(0,1 / 2) & \text { for even } n_{c}  \label{I.21}\\ (-1,-1 / 2) \cup(1 / 2,1) & \text { for odd } n_{c}\end{cases}    
\end{align}
$Q$ is neither an integer nor half-integer, and is not necessarily rational. Specific values of $b$ give rational values for $Q$ (fractional topological charge). For example,

\begin{align}
|b| & =\frac{2 \sqrt{n-1}}{n-2} \Rightarrow \\
Q & = \pm \frac{1}{2}\left[1-(-1)^{n_{c}} \frac{n-2}{n}\right] \\
& = \pm \begin{cases}1 / n & \text { for even } n_{c}  \label{I.22}\\ (n-1) / n & \text { for odd } n_{c}\end{cases}
\end{align}

The functions in this group are necessarily exponential asymptotically:
\begin{equation}
\frac{1}{f} \frac{\partial f}{\partial r} \sim b \quad \Rightarrow \quad f(r) \sim e^{b r} \label{I.23}
\end{equation}

From  the definition of $\mathbf{n}(r, \theta)$ in Eq. \eqref{I.11} a uniform limit for $r \rightarrow \infty$ requires the x and y component to go to zero, which is the case in (a.1), but not (a.2) and (a.3). In other words, compactification of the domain $\mathbb{R}^{2}$ is not possible in (a.2) and (a.3). This agrees with the fact that $Q$ is an integer only in case (a.1). We also note that category (a.2) corresponds to the case of so-called merons (half-integer skyrmion number) with the z-component of the pseudo-spin vector being zero at infinity.

\begin{figure}[h]
      \includegraphics[width=0.48 \textwidth]{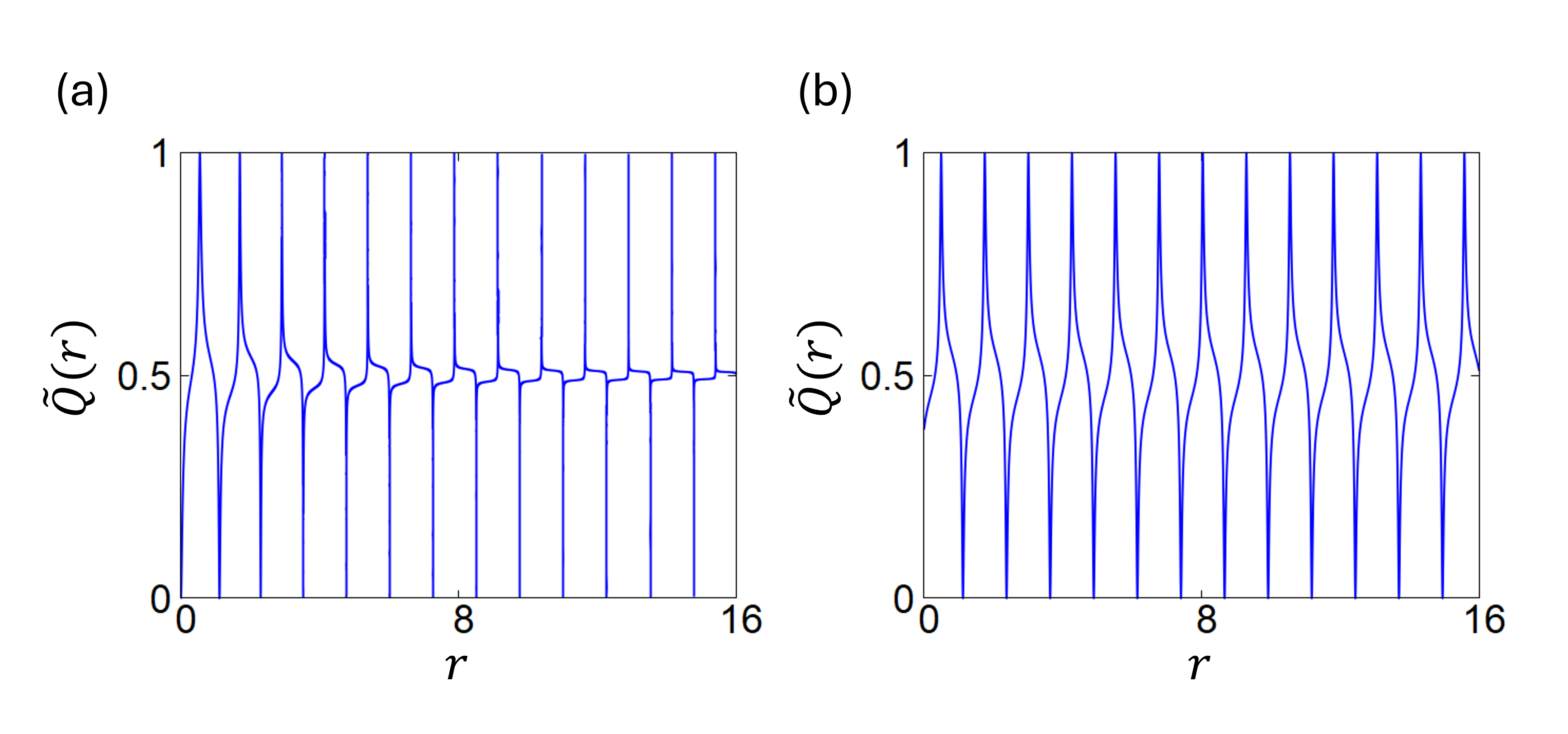} 
    \caption{%
        Schematic of skyrmion integral vs upper integral limit $r$. (a) shows an example from category (b.2) with unitless $r$ and $k=5$ and $\alpha =2$; (b) shows category (b.3) with $k=5$ and $b=-4$. The function (b.3) has a periodicity of $k/2 \pi$, (b.2) is not periodic. Note that both graphs do not decay toward infinity. 
        }
    \label{fig:plot-b2-b3}
\end{figure}

\underline{\textbf{(b)} $f(r)$ crosses zero an infinite number of times.} [And we assume that there does not exist a radius $r_{0}$ such that $f(r)$ does not change sign for $r>r_{0}$.]

Since $f(r) /|f(r)|$ does not tend to a limit as $r \rightarrow \infty, \lim _{r \rightarrow \infty} n_{z}(r)$ does not exist and so the global skyrmion number $Q$ is not well-defined for this function group. The ratio $\frac{1}{f} \frac{\partial f}{\partial r}$ may also fail to converge.

We illustrate the above argument about the non-existence of $Q$ with some examples. We consider functions which are products of trigonometric oscillating functions and the asymptotically decaying functions in group (a) above:
\begin{equation}
f(r) \sim \cos (k r) f_{0}(r) \quad \text { as } \quad r \rightarrow \infty, \quad n_{z}(0)=1 \label{I.24}
\end{equation}
implying
\begin{equation}
\frac{1}{f} \frac{\partial f}{\partial r} \sim-k \tan (k r)+\frac{1}{f_{0}} \frac{\partial f_{0}}{\partial r} \label{I.25}
\end{equation}

\textbf{(b.1)} $\frac{1}{f_{0}} \frac{\partial f_{0}}{\partial r} \rightarrow 0$ when $r \rightarrow \infty$ (e.g. power-law decay).

In this case,
\begin{equation}
n_{z}(r) \sim \frac{f(r)}{|f(r)|} \frac{1}{\sqrt{1+k^{2} \tan ^{2}(k r)}} \label{I.26}
\end{equation}
\begin{equation}
\Rightarrow \quad \tilde{Q}(r) \sim \frac{1}{2}\left[1-\frac{f(r)}{|f(r)|} \frac{1}{\sqrt{1+k^{2} \tan ^{2}(k r)}}\right] \label{I.27}
\end{equation}

The skyrmion integral $\widetilde{Q}(r)$ oscillates between 0 and 1 asymptotically and does not converge.

An example of this case is our recent paper on spontaneous pattern formation in microcavities with a 
Gaussian shaped pump beam \cite{wingenbach-etal.2026skyrmions}.

\textbf{(b.2)} $\frac{1}{f_{0}} \frac{\partial f_{0}}{\partial r} \rightarrow \infty$ 
 when $r \rightarrow \infty$ (e.g. Gaussian).

Using the Gaussian as an example, we have
\begin{eqnarray}
& f_{0}(r) \sim e^{-a r^{2}} \\
& \Rightarrow \quad \frac{1}{f} \frac{\partial f}{\partial r} \sim-k \tan (k r)-2 a r  \label{I.28}\\
& \Rightarrow \quad \tilde{Q}(r) \sim \frac{1}{2}\left[1-\frac{f(r)}{|f(r)|} \frac{1}{\sqrt{1+[k \tan (k r)+2 a r]^{2}}}\right] \label{I.29}
\end{eqnarray}

$\tilde{Q}(r)$ is again oscillating, but it's worth noting that as $r \rightarrow \infty, \tilde{Q}(r) \sim \frac{1}{2}$ except within short intervals around $k r=3 \pi / 2+2 m \pi, m \in \mathbb{Z}$, and the width of these intervals shrinks to zero as $r \rightarrow \infty$.

Particular examples for this case include gain guiding in microchip lasers or Raman amplifiers including two transverse dimensions \cite{druten-etal.2001}.

\textbf{(b.3)}  $\frac{1}{f_{0}} \frac{\partial f_{0}}{\partial r} \rightarrow b$ when $r \rightarrow \infty$ (exponential).\\
In this case,
\begin{align}
f_{0}(r) & \sim e^{b r} \\
& \Rightarrow \frac{1}{f} \frac{\partial f}{\partial r} \sim-k \tan (k r)+b \label{I.30}
\end{align}
\begin{equation}
\Rightarrow \quad \tilde{Q}(r) \sim \frac{1}{2}\left[1-\frac{f(r)}{|f(r)|} \frac{1}{\sqrt{1+[-k \tan (k r)+b]^{2}}}\right] \label{I.31}
\end{equation}

$\tilde{Q}(r)$ oscillates between 0 and 1 asymptotically and does not converge.

\begin{figure}[h]
   \includegraphics[width=0.48 \textwidth]{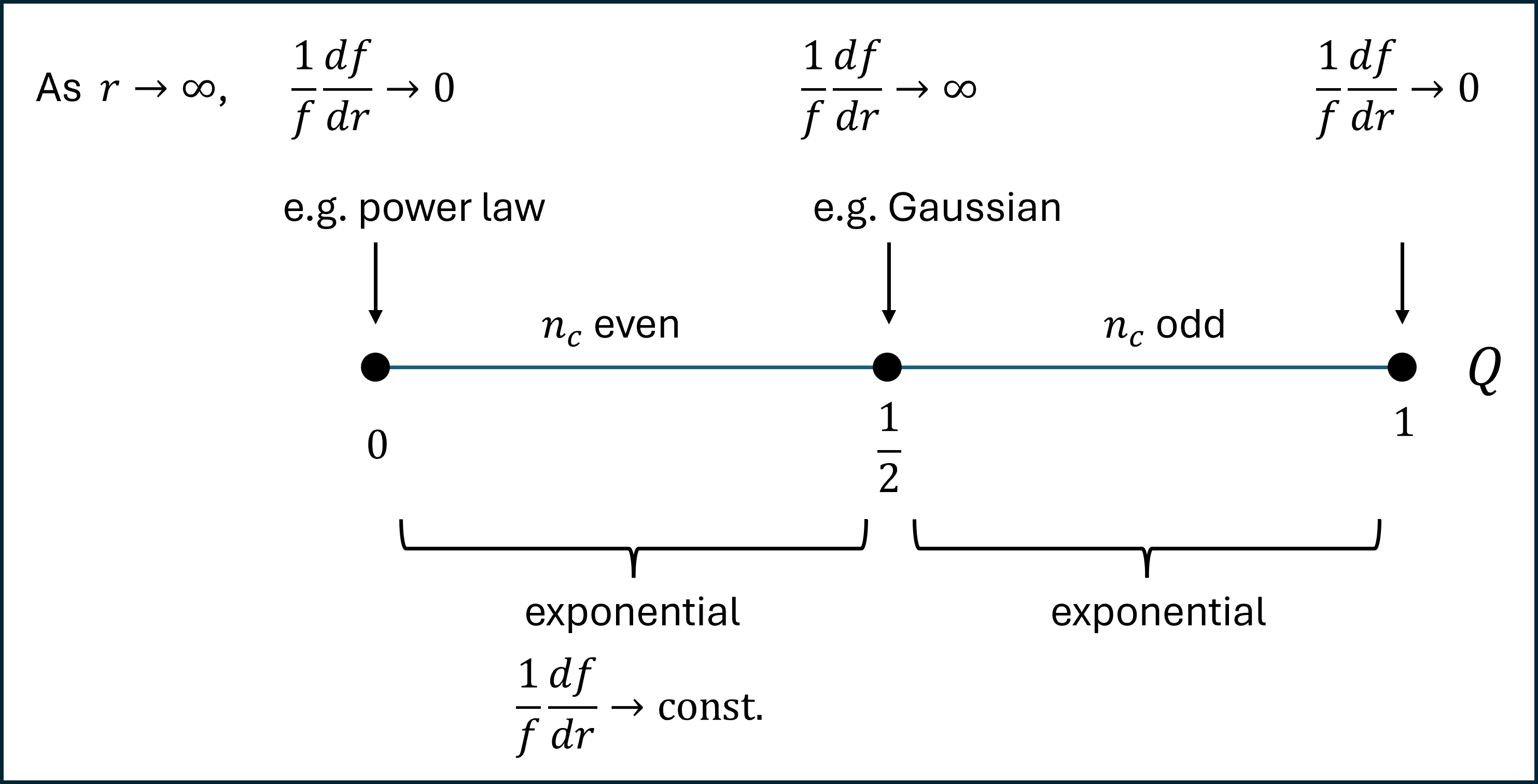} 
    \caption{%
        Schematic summary of the skyrmion number for different behaviors of the function $f$. $n_z(r=0)=+1$ is assumed here (see Eq. \eqref{I.14}); for $n_z(r=0)=-1$ the value of $Q$ changes sign.
        }
    \label{fig:plot-summary_Q}
\end{figure}

The categories (b.2) and (b.3) are illustrated in Fig. \ref{fig:plot-b2-b3}. 
 A schematic summary of the various cases is shown in Fig. \ref{fig:plot-summary_Q}.

\subsubsection{Finite-size disk-like domain}

In this section, we briefly comment on the case of a disk-shaped domain with finite radius $r_0$. In such a domain, we can either have a finite number of zero crossings, or an infinite number, and we restrict the discussion the case of a finite number of zero crossings. The previous analysis leading up to categories (a.1), (a.2) and (a.3) remains valid if the upper integral limit is taken to be finite, $r \rightarrow r_0$, rather than the limit  $r \rightarrow \infty$.

Analogous to category (a.1), if $\frac{1}{f} \frac{\partial f}{\partial r} \rightarrow 0$ when $r \rightarrow r_0$, the skyrmion number is given by Eq. \eqref{I.16}.
Similarly, similar to  category (a.2), if $\left|\frac{1}{f} \frac{\partial f}{\partial r}\right| \rightarrow \infty$ when $r \rightarrow r_0$, the skyrmion number is given by Eq. \eqref{I.18},
and in analogy to category (a.3), if $\frac{1}{f} \frac{\partial f}{\partial r} \rightarrow b$ when $r \rightarrow r_0$ where $b$ is a non-zero constant, 
the skyrmion number is given by Eq. \eqref{I.21}.

\subsection{Scalar-field based momentum-space skyrmions}
\label{Sec:Scalar-field-based-momentum-space-skyrmions}

The previous section shows that if the function $f(r)$ keeps oscillating as $r \rightarrow \infty$, even when the oscillations are damped, the unit vector field as defined in Eq. \eqref{I.11} does not yield a well-defined global skyrmion number $Q$. Nevertheless, we can still construct some topological structure out of these fields by analyzing them in Fourier (momentum) space. Why this may work is because an oscillating function in configuration space ($r$ space) tends to have a Fourier-transformed ($k$ space) counterpart that behaves as a power law asymptotically at large $k$ and has only a finite number of zero crossings over $k \in(0, \infty)$. We illustrate this point with the Hankel function $H_{0}^{(1)}$.

We note that, quite generally, Hankel functions (Bessel functions of the third kind) and other
Bessel functions are important in optical wave propagation (specifically solutions of the Helmholtz equation)
with cylindrical symmetry \cite{gomez-etal.2017}, and in the context of skyrmions they have recently been discussed, for example, in Refs. 
\cite{singh-etal.2023,neuhaus-etal-1.2026arxiv,neuhaus-etal-2.2026arxiv}.

The Hankel function $H_{0}^{(1)}(\alpha r)$ with a complex $\alpha=\alpha^{\prime}+i \alpha^{\prime \prime}$ where $\alpha^{\prime}, \alpha^{\prime \prime} \in \mathbb{R}$, represents a 2D damped, out-going wave $\left(\alpha^{\prime}, \alpha^{\prime \prime}>0\right)$. It does not have a well-defined global skyrmion number through the vector field defined in Eq. \eqref{I.11}. The Fourier transform of $H_{0}^{(1)}(\alpha r)$ is
\begin{equation}
\widetilde{H}_{0}^{(1)}(k, \alpha)=\int d^{2} r e^{-i \vec{k} \cdot \vec{r}} H_{0}^{(1)}(\alpha r)=-\frac{4 i}{k^{2}-\alpha^{2}} \label{II.1}
\end{equation}

The integral in Eq. \eqref{II.1} is well-defined only for $\alpha^{\prime \prime}>0$. Although the result in Eq. \eqref{II.1} is known, we include a derivation in Appendix \ref{Sec:FT-of-Hankel} for completeness. If $k$ is extended from the positive real axis to the complex plane, $\widetilde{H}_{0}^{(1)}(k, \alpha)$ as given has two poles $k_{\text {pole }}= \pm \alpha$. The two poles approach the real axis in the weak-damping 
($\alpha^{\prime \prime} \downarrow 0$) limit, implying, for $k, \alpha^{\prime}>0$,
\begin{equation}
\lim _{\alpha^{\prime \prime} \downarrow 0} \widetilde{H}_{0}^{(1)}(k, \alpha)=2 \pi\left[\frac{1}{k} \delta\left(k-\alpha^{\prime}\right)-\frac{2}{\pi} \wp \frac{1}{k^{2}-\alpha^{\prime 2}} i\right] \label{II.2}
\end{equation}
[see Appendix, \ref{Sec:FT-of-Hankel}] where $\wp$ denotes the principal value integral.

\begin{figure}[t]
     \includegraphics[width=0.48 \textwidth]{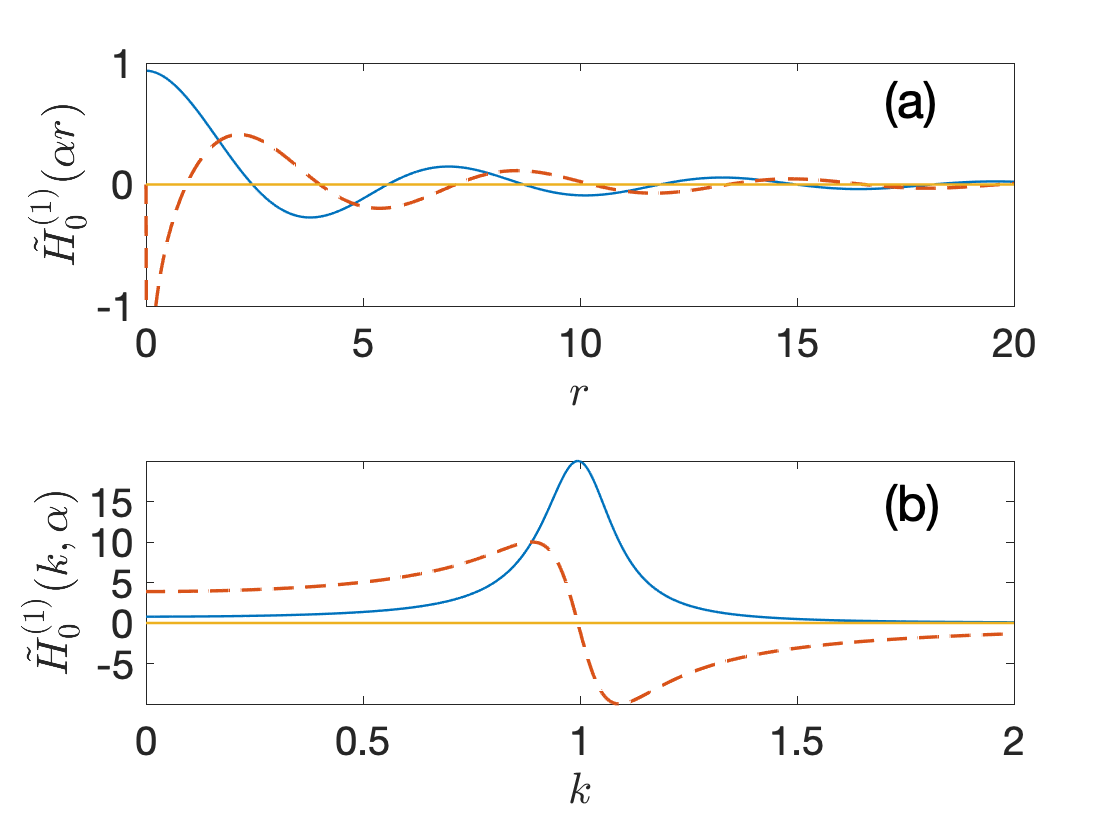} 
    \caption{%
        Example of Hankel function in real space (a) and its Fourier transform in momentum space (b). The coordinates $r$ and $k$ are taken to be unitless, and the value of $\alpha=1+0.1i$ is chosen. The blue solid line (red dashed line)  shows the real (imaginary) part. In (a) we have infinitely many zero crossings, while in (b) we have zero (one) zero crossing for the real (imaginary) part.
        }
    \label{fig:plot-plot-Hankel}
\end{figure}

For convenience, we show in Fig. \ref{fig:plot-plot-Hankel} a Hankel function with complex argument in real space and in momentum space. We see the usual infinitely continued oscillation in real space, in which the number of zero-crossings $n_c$ is infinite, and the corresponding momentum-space behavior, in which $n_c =0$   ($n_c =1$)   for the real  (imaginary) part of the Fourier-transformed Hankel function.

To construct a skyrmionic vector field  $\mathbf{n}_{1}(\mathbf{k}, \alpha)$   from the scalar function $\widetilde{H}_{0}^{(1)}(k, \alpha)$, we choose the real-valued scalar function $f(k)$ to be either $\operatorname{Re} \widetilde{H}_{0}^{(1)}(k, \alpha)$ or $\operatorname{Im} \widetilde{H}_{0}^{(1)}(k, \alpha)$. In a similar manner as in the treatment in $r$--space, the vector field becomes
\begin{eqnarray}
\mathbf{n}_{1}(\mathbf{k}, \alpha)
 &=\frac{1}{\sqrt{f^{2}+\left(\frac{\partial f}{\partial k}\right)^{2}}}\left[\begin{array}{c}
\frac{\partial f}{\partial k_{x}}  \label{II.3}\\
\frac{\partial f}{\partial k_{y}} \\
f
\end{array}\right] 
\\
&=\frac{1}{\sqrt{f^{2}+\left(\frac{\partial f}{\partial k}\right)^{2}}}\left[\begin{array}{c}
\cos \theta_{k} \frac{\partial f}{\partial k} \\
\sin \theta_{k} \frac{\partial f}{\partial k} \\
f
\end{array}\right]
\end{eqnarray}
where $\mathbf{k}=\left(k_{x}, k_{y}\right)=\left(k, \theta_{k}\right)$. Taking the derivative of $\widetilde{H}_{0}^{(1)}(k, \alpha)$ given above, we have
\begin{equation}
\frac{\partial \widetilde{H}_{0}^{(1)}(k, \alpha)}{\partial k}=\frac{8 k i}{\left(k^{2}-\alpha^{2}\right)^{2}} \label{II.4}
\end{equation}

For the development below, we expand the expressions for $\widetilde{H}_{0}^{(1)}(k, \alpha)$ and $\frac{\partial \widetilde{H}_{0}^{(1)}(k, \alpha)}{\partial k}$ into their real and imaginary parts explicitly:
\begin{eqnarray}
\widetilde{H}_{0}^{(1)}(k, \alpha) &=&\frac{8 \alpha^{\prime} \alpha^{\prime \prime}-4 \zeta i}{\zeta^{2}+4 \alpha^{\prime 2} \alpha^{\prime \prime 2}}  \\
 \quad \frac{\partial \widetilde{H}_{0}^{(1)}(k, \alpha)}{\partial k} &=& \frac{-32 k \alpha^{\prime} \alpha^{\prime \prime} \zeta+8 k\left[\zeta^{2}-4 \alpha^{\prime 2} \alpha^{\prime \prime 2}\right] i}{\left[\zeta^{2}+4 \alpha^{\prime 2} \alpha^{\prime \prime 2}\right]^{2}}  \nonumber \\ \label{II.5}
\end{eqnarray}
where $\zeta=k^{2}-\alpha^{\prime 2}+\alpha^{\prime \prime 2}$.\\
In parallel with the skyrmion numbers in configuration space, we define the momentum space skyrmion integral and global skyrmion number as
\begin{align}
    \tilde{Q}_{1}(K) & =\frac{1}{4 \pi} \int_{0}^{K} d k k \int_{0}^{2 \pi} d \theta_{k} \mathbf{n}_{1} \cdot\left[\frac{\partial \mathbf{n}_{1}}{\partial k_{x}} \times \frac{\partial \mathbf{n}_{1} } {\partial k_y}\right]  \label{II.6}
\end{align}
\begin{align}
Q_{1} & =\lim _{K \rightarrow \infty} \tilde{Q}_{1}(K)  \label{II.7a} \\
&=\frac{1}{4 \pi} \int_{0}^{\infty} d k k \int_{0}^{2 \pi} d \theta_{k} \mathbf{n}_{1}  \cdot\left[\frac{\partial \mathbf{n}_{1} }{\partial k_{x}} \times \frac{\partial \mathbf{n}_{1}   }{\partial k_y}\right] \label{II.7}
\end{align}

Following the same algebraic procedure as that in the treatment in $r$--space, we reduce $\tilde{Q}_{1}(K)$ and $Q_{1}$ in Eqs. \eqref{II.6}  and \eqref{II.7} to
\begin{align}
\tilde{Q}_{1}(K)=\frac{1}{2}\left[n_{1 z}(0)-n_{1 z}(K)\right]
\end{align}
\begin{align}
 Q_{1} & =\lim _{K \rightarrow \infty} \tilde{Q}_{1}(K) \\
 & =\frac{1}{2}\left[n_{1 z}(0)-\lim _{K \rightarrow \infty} n_{1 z}(K)\right] \label{II.8}
\end{align}

We will now analyze the momentum space skyrmion numbers in the two cases where the function $f(k)$ is given by the real part (A) or the imaginary part (B) of the Fourier-transformed Hankel function.

First, for case (A), $f(k)=\operatorname{Re} \widetilde{H}_{0}^{(1)}(k, \alpha) \equiv f_{R}(k)$. We have, from Eq. \eqref{II.5},
\begin{align}
f_{R}(k) &=\frac{8 \alpha^{\prime} \alpha^{\prime \prime}}{\zeta^{2}(k)+4 \alpha^{\prime 2} \alpha^{\prime \prime 2}}  \\
 \frac{\partial f_{R}}{\partial k} & =-\frac{32 k \alpha^{\prime} \alpha^{\prime \prime} \zeta(k)}{\left[\zeta^{2}(k)+4 \alpha^{\prime 2} \alpha^{\prime \prime 2}\right]^{2}} \label{II.9}
\end{align}
leading to
\begin{equation}
\frac{1}{f_{R}} \frac{\partial f_{R}}{\partial k}=-\frac{4 k \zeta(k)}{\zeta^{2}(k)+4 \alpha^{\prime 2} \alpha^{\prime \prime 2}} \label{II.10}
\end{equation}

The $z$ component of the vector field, Eq. \eqref{II.3} becomes
\begin{align}
n_{1z}(k, \alpha)
 & =\frac{f_{R}}{\sqrt{f_{R}^{2}+\left(\frac{\partial f_{R}}{\partial k}\right)^{2}}} \\
 &=\frac{f_{R}}{\left|f_{R}\right|} \frac{1}{\sqrt{1+\left(\frac{1}{f_{R}} \frac{\partial f_{R}}{\partial k}\right)^{2}}} \\
& =\operatorname{sign}\left[\alpha^{\prime} \alpha^{\prime \prime}\right] \frac{1}{\sqrt{1+\left(\frac{4 k \zeta(k)}{\zeta^{2}(k)+4 \alpha^{\prime 2} \alpha^{\prime \prime 2}}\right)^{2}}} \label{II.11}
\end{align}

For the damped, out-going waves we are considering, $\operatorname{sign}\left[\alpha^{\prime} \alpha^{\prime \prime}\right]=1$. At $k=0$,
\begin{equation}
\frac{1}{f_{R}} \frac{\partial f_{R}}{\partial k}=0 \quad \Longrightarrow \quad n_{1 z}(0, \alpha)=1 \label{II.12}
\end{equation}

Giving the skyrmion integral at a finite upper radius $K$ as
\begin{equation}
\tilde{Q}_{1}(K)=\frac{1}{2}\left(1-\frac{1}{\sqrt{1+\left(\frac{4 K \zeta(K)}{\zeta^{2}(K)+4 \alpha^{\prime 2} \alpha^{\prime \prime 2}}\right)^{2}}}\right) \label{II.13}
\end{equation}

As $K \rightarrow \infty$,
\begin{align}
\zeta(K) \sim K^{2} \Rightarrow f_{R}(K) \sim \frac{8 \alpha^{\prime} \alpha^{\prime \prime}}{K^{4}} \left. 
 \frac{\partial f_{R}}{\partial k}\right|_{k=K} 
 \end{align}
 \begin{align}
   &=-\frac{32 \alpha^{\prime} \alpha^{\prime \prime}}{K^{5}}  \Rightarrow \frac{1}{f_{R}(K)} \left. \frac{\partial f_{R}}{\partial k}\right|_{k=K} \sim-\frac{4}{K} \label{II.14}
\end{align}
implying
\begin{equation}
n_{1 z}(K, \alpha) \sim \frac{1}{\sqrt{1+\left(\frac{4}{K}\right)^{2}}} \sim 1 \label{II.15}
\end{equation}

So the global skyrmion number, denoted $Q_{R}$, vanishes:
\begin{equation}
Q_{R}=\frac{1}{2}\left[n_{1 z}(0, \alpha)-\lim _{K \rightarrow \infty} n_{1 z}(K, \alpha)\right]=0 \label{II.16}
\end{equation}

We now turn to case (B) where $f(k)=\operatorname{Im} \widetilde{H}_{0}^{(1)}(k, \alpha) \equiv f_{I}(k)$. From Eq. \eqref{II.5} we have
\begin{align}
f_{I}(k)&=-\frac{4 \zeta(k)}{\zeta^{2}(k)+4 \alpha^{\prime 2} \alpha^{\prime \prime 2}} \\
\frac{\partial f_{I}}{\partial k}&=\frac{8 k\left[\zeta^{2}(k)-4 \alpha^{\prime 2} \alpha^{\prime \prime 2}\right]}{\left[\zeta^{2}(k)+4 \alpha^{\prime 2} \alpha^{\prime \prime 2}\right]^{2}} \label{II.17}
\end{align}
leading to
\begin{equation}
\frac{1}{f_{I}} \frac{\partial f_{I}}{\partial k}=-\frac{2 k\left[\zeta^{2}(k)-4{\alpha^{\prime}}^{2} \alpha^{\prime \prime 2}\right]}{\zeta(k)\left[\zeta^{2}(k)+4{\alpha^{\prime 2}}^{\prime 2} \alpha^{\prime \prime 2}\right]} \label{II.18}
\end{equation}

We now need to distinguish between the cases where $\alpha^{\prime}=\alpha^{\prime \prime}$ and $\alpha^{\prime} \neq \alpha^{\prime \prime}$. If $\alpha^{\prime}=\alpha^{\prime \prime}$ and therefore $\zeta(k)=k^{2}$, we have
\begin{align}
f_{I}(k)&=-\frac{4 k^{2}}{k^{4}+4 \alpha^{\prime 4}} \\
 \frac{\partial f_{I}}{\partial k}&=\frac{8 k\left[k^{4}-4 \alpha^{\prime 4}\right]}{\left[k^{4}+4 \alpha^{\prime 4}\right]^{2}} \label{II.19}
\end{align}
and
\begin{equation}
\frac{1}{f_{I}} \frac{\partial f_{I}}{\partial k}=-\frac{2\left[k^{4}-4 \alpha^{\prime 4}\right]}{k\left[k^{4}+4 \alpha^{\prime 4}\right]} \label{II.20}
\end{equation}

The sign of $f_{I}(k)$ being negative for all $k$, the $z$ component of the vector field is
\begin{align}
n_{1 z}(k, \alpha)& =\frac{f_{I}}{\left|f_{I}\right|} \frac{1}{\sqrt{1+\left(\frac{1}{f_{I}} \frac{\partial f_{I}}{\partial k}\right)^{2}}} \\
&=-\frac{1}{\sqrt{1+\frac{4}{k^{2}}\left(\frac{k^{4}-4 \alpha^{\prime 4}}{k^{4}+4 \alpha^{\prime 4}}\right)^{2}}} \label{II.21}
\end{align}
As $k \downarrow 0$,
\begin{align}
f_{I}(k) & \sim-\frac{k^{2}}{\alpha^{\prime 4}} \\
\frac{\partial f_{I}}{\partial k}&  \sim-\frac{2 k}{\alpha^{\prime 4}} \\
 \frac{1}{f_{I}} \frac{\partial f_{I}}{\partial k} &\sim \frac{2}{k} \\
  n_{1 z}(k, \alpha) &\sim-\frac{k}{2} \label{II.22}
\end{align}

The vanishing of $n_{1 z}(0)$ implies that the in-plane components of the skyrmion field vector assume a vortex structure with a non-vanishing magnitude around $k=0: \lim _{k \downarrow 0} n_{1 x}(k)=$ $\cos \theta_{k}, \lim _{k \downarrow 0} n_{1 y}(k)=\sin \theta_{k}$. Hence the skyrmion field vector $\mathbf{n}_{1}(k)$ is not well-defined at $k=0$. With this caveat, we can nevertheless proceed formally and obtain the skyrmion integral at a finite upper limit $K$ as
\begin{equation}
\tilde{Q}_{1}(K)=\frac{1}{2 \sqrt{1+\frac{4}{K^{2}}\left(\frac{K^{4}-4{\alpha^{\prime}}^{4}}{K^{4}+4{\alpha^{\prime}}^{4}}\right)^{2}}} \label{II.23}
\end{equation}

The global skyrmion number is given by
\begin{equation}
Q_{1}=\lim _{K \rightarrow \infty} \tilde{Q}_{1}(K)=\frac{1}{2} \label{II.24}
\end{equation}

We note that this value of $Q_{1}$ does not contradict the result for group (a.1) in the previous section that the global skyrmion number derived from a power-law decay function is an integer because the functions considered in that section are assumed to be finite at the origin ($f(0) \neq$ $0)$. This condition is not satisfied here ($f_{I}(0)=0$).

The half-integer global skyrmion number corresponds to a meron (here in momentum space) where 
the pseudo-spin vector field $\mathbf{n}_1(\mathbf{k})$  covers half the $S^2$ target space.
Commonly, merons are spin (or pseudo-spin) structures that are well defined at the origin (of the physical space, which, again, is the momentum space in our present discussion), and that form vortex structures, either purely radial or purely azimuthal, at the periphery of the physical space (which in our case is the limit $k \rightarrow \infty$). In our case, the opposite is true, namely that the pseudo-spin texture forms a vortex in the vicinity of the origin, and is well defined with a uniform limit when $k \rightarrow \infty$. In the case of the periphery at infinity, the conventional meron does not allow for compactification because there is no uniform limit of the spin at infinity (it depends on the direction in which infinity is approached). Similarly, our case does not allow for compactification, because there is no uniform limit at the origin  ($k \rightarrow 0$).

When $\alpha^{\prime} \neq \alpha^{\prime \prime}$, we have, from Eqs. \eqref{II.17} and \eqref{II.18},
\begin{align}
n_{1 z}(k, \alpha) & =\frac{f_{I}}{\left|f_{I}\right|} \frac{1}{\sqrt{1+\left(\frac{1}{f_{I}} \frac{\partial f_{I}}{\partial k}\right)^{2}}} \\
&=-\frac{\operatorname{sign}[\zeta(k)]}{\sqrt{1+\frac{4 k^{2}}{\zeta^{2}(k)}\left(\frac{\zeta^{2}(k)-4{\alpha^{\prime}}^{2} \alpha^{\prime \prime 2}}{\zeta^{2}(k)+4{\alpha^{\prime}}^{2} \alpha^{\prime \prime 2}}\right)^{2}}} \label{II.25}
\end{align}

In the limit $k \downarrow 0$,
\begin{align}
 & f_{I}(0)=-\frac{4\left(\alpha^{\prime 2}-\alpha^{\prime \prime 2}\right)}{|\alpha|^{4}}  \\
 & \left.\quad \frac{\partial f_{I}}{\partial k}\right|_{k=0} \sim \frac{8 k\left[\left(\alpha^{\prime 2}-\alpha^{\prime \prime 2}\right)^{2}-4{\alpha^{\prime 2}}^{\prime \prime 2}\right]}{|\alpha|^{8}} \rightarrow 0 \label{II.26}
\end{align}
and therefore
\begin{equation}
\left.\frac{1}{f_{I}(0)} \frac{\partial f_{I}}{\partial k}\right|_{k=0}=0 \label{II.27}
\end{equation}
and
\begin{equation}
n_{1 z}(0, \alpha)=\operatorname{sign}\left({\alpha^{\prime}}^{2}-{\alpha^{\prime \prime}}^{2}\right) \label{II.28}
\end{equation}

At a finite upper limit $K$, the skyrmion integral is given by
\begin{widetext}
\begin{equation}
\tilde{Q}_{1}(K)=
\frac{1}{2}
\left[
\operatorname{sign}\left({\alpha^{\prime}}^{2}-{\alpha^{\prime \prime}}^{2}\right)+\frac{\operatorname{sign}[\zeta(K)]}
{\sqrt{1+
\frac{4 K^{2}}{\zeta^{2}(K)}
\left(
\frac{\zeta^{2}(K)-4{\alpha^{\prime 2} \alpha^{\prime \prime 2}}}
{\zeta^{2}(K)+4{\alpha^{\prime 2} \alpha^{\prime \prime 2}}}
\right)^2 
}
}
\right]
 \label{II.29}
\end{equation}
\end{widetext}

At the $k \rightarrow \infty$ limit,
\begin{align}
& f_{I}(k)  \sim-\frac{4}{k^{2}} \\
& \frac{\partial f_{I}}{\partial k} \sim \frac{8}{k^{3}} \\
& \frac{1}{f_{I}} \frac{\partial f_{I}}{\partial k} \sim-\frac{2}{k} \rightarrow 0 \label{II.30}
\end{align}
and thus
\begin{equation}
\lim _{k \rightarrow \infty} n_{1 z}(k, \alpha)=-1 \label{II.31}
\end{equation}
and
\begin{align}
Q_{1} &=\lim _{K \rightarrow \infty} \tilde{Q}_{1}(K)  =\frac{1}{2}\left[\operatorname{sign}\left(\alpha^{\prime 2}-\alpha^{\prime \prime 2}\right)+1\right]  \label{II.32}\\
& =\left\{\begin{array}{rrr}
1 & \text { if } \alpha^{\prime 2}>\alpha^{\prime \prime 2} & \text { (underdamped) } \\
0 & \text { if } \alpha^{\prime 2}<\alpha^{\prime \prime 2} & \text { (overdamped) }
\end{array}\right. \label{II.33}
\end{align}

It can be seen in Eq. \eqref{II.25}  that $n_{1 z}(K, \alpha)$ 
changes sign
at $\zeta(K)=0$. If $\alpha^{\prime 2}>\alpha^{\prime \prime 2}$, this happens at one value of $K$ :
\begin{eqnarray}
\operatorname{sign}[\zeta(K)]=\left\{\begin{array}{cl}
-1 & \text { if } K^{2}<\alpha^{\prime 2}-\alpha^{\prime \prime 2}  \label{II.34}\\
1 & \text { if } K^{2}>\alpha^{\prime 2}-\alpha^{\prime \prime 2}
\end{array}\right.
\end{eqnarray}
and therefore
\begin{eqnarray}
-n_{1 z}(K, \alpha) \begin{cases}<0 & \text { if } K^{2}<\alpha^{\prime 2}-\alpha^{\prime \prime 2}  \label{II.35}\\ >0 & \text { if } K^{2}>\alpha^{\prime 2}-\alpha^{\prime \prime 2}\end{cases}
\end{eqnarray}

Conversely, if $\alpha^{\prime 2}<\alpha^{\prime \prime 2}$, then $\operatorname{sign}[\zeta(K)]=1$, and
\begin{equation}
-n_{1 z}(K, \alpha)>0 \quad \text { for all } K \label{II.36}
\end{equation}

The skyrmions we consider in this paper are maps from the $\mathbb{R}^{2}$ base space (2D real or momentum space) to an $S^{2}$ target space (3--component, unit-norm, real-valued vectors in $\mathbb{R}^{3}$ ). For this class of field configurations, there exists a map
 between the target space and the state space of a quantum mechanical two-level system,
for example Refs. \cite{nakahara.2003,asboth-etal.2016,vanderbilt.2018}.
This map allows us to use the concepts of Berry connection and Berry phase, defined via the quantum state vectors, in our discussion of the skyrmions. Appendix  \ref{Sec:appendix-relation-chern} summarizes one way to set up the map. A $2 \times 2$ matrix Hamiltonian in the quantum domain is constructed as the scalar product of the skyrmion field vector and the three Pauli matrices treated as a vector. In this way, the Hamiltonian and its eigenvectors can be considered as being parameterized, through the skyrmion field, by the configuration space or momentum space coordinates. With this parameterization, the Berry connection and Berry curvature are derived in the usual manner. The skyrmion density and the corresponding Berry curvature are equal, and, when Stokes theorem applies, the skyrmion integral over a finite, simply-connected domain is equal to the Berry phase along the closed-loop boundary divided by $2 \pi$ (the gauge in the Berry phase is chosen for the equality to be satisfied). In cases where the global skyrmion number exists, and the $\mathbb{R}^{2}$ base space can be compactified to $S^{2}$, as in the case with $f$ having a power-law decay
and a finite number of zero crossings,
the global skyrmion number is equal to the first Chern number
(for example Refs. \cite{hasan-kane.2010,fruchart-carpentier.2013,cook.2023}).

We finally note that  two recent papers, Ref. \cite{neuhaus-etal-1.2026arxiv} and Ref. \cite{neuhaus-etal-2.2026arxiv}, define a momentum space topological invariant, called the momentum space linking number (MoLiN), that characterizes wave fields in 2D space, including those whose real space representations fail to yield well-defined global skyrmion numbers. In their framework, if the Fourier transform of a wave field is localized around a ring (or a closed loop homotopic to a ring) in momentum space, a Berry phase can be constructed from the field along the ring. The winding number of the Berry phase gives the MoLiN \cite{neuhaus-etal-1.2026arxiv}.

There are differences between the topological information carried by our momentum space global skyrmion number and that carried by the MoliN. As noted before, the base space of the momentum space skyrmion is $\mathbb{R}^{2}$ (in some cases compactified to $S^{2}$ ) and the target space is $S^{2}$, whereas both the base and the target spaces of the MoLiN are loops ($S^{1}$). For the skyrmions considered here, for which the defining scalar function $f$ is circularly symmetric, the global skyrmion number is determined, as shown above, by the functional behavior of $f(k)$ at large radial $k=|\mathbf{k}|$. In contrast, the MoLiN does not have any crucial dependence on the functional behavior of the field along the $k$ radius apart from the field's narrow localization around a loop. Angular variation of the field is the main contributor to the MoLiN. 
While our skyrmions have vorticity $\pm 1$ (Eq. (5)), the polariton fields underpinning them have zero angular momentum. The MoLiN equals zero for  fields with zero angular momentum \cite{neuhaus-etal-1.2026arxiv} \cite{neuhaus-etal-2.2026arxiv}. The fact that the skyrmion numbers of our momentum-space skyrmions can be non-zero shows that the skyrmion number and the MoLiN are two different topological quantities.

We note that the assumption of circular symmetry is made only for the sake of analytical tractability. In general, the scalar function $f(\mathbf{k})$ can of course also have angular dependency, in which case the skyrmion number may receive contributions from components with non-zero angular momentum. How these contributions are related to the MoLiN 
will be addressed in  future research.
We also note that oscillating functions in real space do not necessarily have Fourier transforms that are narrowly peaked in momentum space. The Fourier transform of the zero-order Hankel function, for example, has an imaginary part that shows a Lorentzian-like behavior even when the dissipation rate approaches zero (see Eqs. 
\eqref{II.1} and \eqref{II.2}).


\section{Conclusion}

\label{Sec:conclusion}
    
We have analyzed and categorized the skyrmion integral for normalized real-valued pseudo-spin vectors formed from real-valued  scalar fields for various functional forms of the scalar field (restricted to zero angular momentum fields). We distinguished skyrmion integrals for skyrmion-like textures in configuration space (or real space) versus those in momentum space. The classes of functions that we analyzed are all based on (or representative of) recent examples found in the literature using optical and water waves. 
We have shown that the asymptotic behavior determines whether the scalar-wave texture has a genuine global topology, and Fourier transformation can produce a well-defined integer momentum-space topology even where no global real-space skyrmion number exists.

In particular, in configuration  space (real space), we distinguish  six categories. 
First, the three categories where the number of zero crossings, $n_c$, is finite,
(a.1)-(a.3). The skyrmion number in  category (a.1) is
integer (specifically, it is $0$ or $\pm 1$), see Eq. \eqref{I.16}, in category (a.2)  half-integer ($\pm 1/2$), see Eq. \eqref{I.18}, and in category (a.3) it is not an integer, see Eq. \eqref{I.21}. 
The next category is where the number of zero crossings is infinite,
(b.1)-(b.3). Here, $\tilde{Q}(r)$ oscillates asymptotically and 
the skyrmion number $Q$ is not defined (the integral does not converge), see Eq. \eqref{I.27} for category (b.1), 
Eq. \eqref{I.29} for category (b.2), and
Eq. \eqref{I.31} for category (b.3). We note, however, that in category (b.2) the skyrmion number $\tilde{Q}(r)$ approaches $1/2$ in the limit $r \rightarrow \infty$  for almost all values of $r$ (see Fig. \ref{fig:plot-b2-b3}(a)).

The situation is very different in momentum space, as - at least in our example of the Hankel function - the scalar field can only have zero or one zero crossing, depending on whether we define the skyrmion number with respect to the real part or the imaginary part of the scalar field, see Fig. \ref{fig:plot-plot-Hankel}. Therefore, we find the skyrmion number in momentum space to be either zero, Eq. \eqref{II.16}, or $1$ or $0$, Eq. \eqref{II.33}, depending on whether the function can be classified as underdamped or overdamped, respectively.

In future work we plan to extend this analysis to a broader range of underlying scalar functions, in particular those with non-zero angular momentum.  Moreover, this work was motivated by our recent study of non-Hermitian microcavities,  and the extension and application of these results to this area is underway.  More generally, we intend to explore the extension to complex valued unit vector fields.

\begin{acknowledgments}
    The authors gratefully acknowledge financial support for the Arizona group from the US National Science Foundation (NSF) under Grant No. DMR-1839570, and, for  the Paderborn group, by the Deutsche Forschungsgemeinschaft (DFG, German Research Foundation) through Grant No.~519608013.\\
\end{acknowledgments}


\appendix

\section{Intermediate steps in skyrmion integral}
\label{Sec:appendix-nagaosa-formula}

In this appendix, we provide, for convenience and completeness, some of the intermediate steps for re-writing the skyrmion number in polar coordinates and then assume a separable form as used in \cite{nagaosa-etal.2013}.

The coordinate transformation of the derivative of the unit vector $\mathbf{n}$ from Cartesian to polar coordinates is
\begin{eqnarray}
{\left[\begin{array}{l}
\frac{\partial \mathbf{n}}{\partial x} \\
\frac{\partial \mathbf{n}}{\partial y}
\end{array}\right]=} & {\left[\begin{array}{ll}
\frac{\partial r}{\partial x} & \frac{\partial \theta}{\partial x} \\
\frac{\partial r}{\partial y} & \frac{\partial \theta}{\partial y}
\end{array}\right]\left[\begin{array}{l}
\frac{\partial \mathbf{n}}{\partial r} \\
\frac{\partial \mathbf{n}}{\partial \theta}
\end{array}\right] } \\
& =\left[\begin{array}{cc}
\cos \theta & -\frac{1}{r} \sin \theta \\
\sin \theta & \frac{1}{r} \cos \theta
\end{array}\right]\left[\begin{array}{l}
\frac{\partial \mathbf{n}}{\partial r} \\
\frac{\partial \mathbf{n}}{\partial \theta}
\end{array}\right] \label{A.1}
\end{eqnarray}
from which we get
\begin{equation}
\frac{\partial \mathbf{n}}{\partial x} \times \frac{\partial \mathbf{n}}{\partial y}=\frac{1}{r}\left(\frac{\partial \mathbf{n}}{\partial r} \times \frac{\partial \mathbf{n}}{\partial \theta}\right) \label{A.2}
\end{equation}
This gives the final form of the skyrmion integral in Eq. \eqref{I.1}
\begin{equation}
\tilde{Q}(r)=\frac{1}{4 \pi} \int_{0}^{r} d r^{\prime} \int_{0}^{2 \pi} d \theta \mathbf{n} \cdot\left[\frac{\partial \mathbf{n}}{\partial r^{\prime}} \times \frac{\partial \mathbf{n}}{\partial \theta}\right] \label{A.3}
\end{equation}

Suppose the unit vector $\mathbf{n}$ is in a separable form
\begin{eqnarray}
\mathbf{n}(r, \theta)=\left[\begin{array}{c}
\cos \Phi(\theta) \sin R(r)  \label{A.4}\\
\sin \Phi(\theta) \sin R(r) \\
\cos R(r)
\end{array}\right]
\end{eqnarray}
Then
\begin{eqnarray}
\frac{\partial \mathbf{n}}{\partial r} \times \frac{\partial \mathbf{n}}{\partial \theta}=\left[\begin{array}{c}
\cos \Phi(\theta) \sin ^{2} R(r)  \label{A5}\\
\sin \Phi(\theta) \sin ^{2} R(r) \\
\cos R(r) \sin R(r)
\end{array}\right] \frac{\partial R}{\partial r} \frac{\partial \Phi}{\partial \theta}
\end{eqnarray}
and
\begin{equation}
\mathbf{n} \cdot\left[\frac{\partial \mathbf{n}}{\partial r} \times \frac{\partial \mathbf{n}}{\partial \theta}\right]=\sin R(r) \frac{\partial R}{\partial r} \frac{\partial \Phi}{\partial \theta} \label{A.6}
\end{equation}

With Eq. \eqref{A.6} as the integrand, the integral in Eq. \eqref{A.3} can be done analytically:
\begin{eqnarray}
& \tilde{Q}(r)=\frac{1}{4 \pi} \int_{0}^{r} d r^{\prime} \sin R\left(r^{\prime}\right) \frac{\partial R}{\partial r^{\prime}} \int_{0}^{2 \pi} d \theta \frac{\partial \Phi}{\partial \theta} \\
&=-\frac{1}{4 \pi} \quad\left[\cos R\left(r^{\prime}\right)\right]_{r^{\prime}=0}^{r^{\prime}=r} \quad[\Phi(\theta)]_{\theta=0}^{\theta=2 \pi} \label{A7}
\end{eqnarray}



\section{Fourier transform of Hankel function }

\label{Sec:FT-of-Hankel}

The 2D Fourier transform of $H_{0}^{(1)}(\alpha r), \alpha \in \mathbb{C}$ is
\begin{widetext}
\begin{align}
\widetilde{H}_{0}^{(1)}(k, \alpha) & =\int_{0}^{\infty} d r r \int_{0}^{2 \pi} d \theta e^{-i k r \cos \theta} H_{0}^{(1)}(\alpha r) \\
& =\int_{0}^{\infty} d r r \sum_{m=-\infty}^{\infty}(-i)^{m} J_{m}(k r) H_{0}^{(1)}(\alpha r) \int_{0}^{2 \pi} d \theta e^{-i m \theta} \\
& =2 \pi \int_{0}^{\infty} d r r J_{0}(k r) H_{0}^{(1)}(\alpha r)  \label{B1}\\
& =2 \pi\left[\frac{\alpha r H_{1}^{(1)}(\alpha r) J_{0}(k r)-k r H_{0}^{(1)}(\alpha r) J_{1}(k r)}{\alpha^{2}-k^{2}}\right]_{r=0}^{r \rightarrow \infty} \label{B2}
\end{align}
\end{widetext}
The expression for the integral in Eq. \eqref{B2} is given on p. 629 of Ref. \cite{gradshteyn-ryzhik.2007}. The asymptotic form of the Hankel functions at large $r$ is given by \cite{abramowitz.72}:

\begin{align}
& H_{v}^{(1)}(\alpha r) \sim \sqrt{\frac{2}{\pi \alpha r}} e^{-\alpha^{\prime \prime} r+i\left(\alpha^{\prime} r-\frac{v \pi}{2}-\frac{\pi}{4}\right)} \nonumber \\
& r \rightarrow \infty, \alpha=\alpha^{\prime}+i \alpha^{\prime \prime}, v \in \mathbb{Z} \label{B3}
\end{align}

We only consider decaying waves, i.e. $\alpha^{\prime \prime}>0$. For this case, since $H_{0}^{(1)}(\alpha r)$ and $H_{1}^{(1)}(\alpha r)$ are decaying exponentially while $J_{0}(k r)$ and $J_{1}(k r)$ remain bounded as $r \rightarrow \infty$, the upper limit of the integral in Eq. \eqref{B2} vanishes. For the lower limit, the behavior of the Bessel functions and Hankel functions near $r=0$ is \cite{abramowitz.72}
\begin{align}
& H_{0}^{(1)}(\alpha r) \sim \frac{2 i}{\pi} \ln (\alpha r) \\
&  H_{1}^{(1)}(\alpha r) \sim-\frac{i}{\pi} \cdot \frac{2}{\alpha r} \\
& J_{0}(k r) \sim 1 \\
& J_{1}(k r) \sim \frac{k r}{2} \\
&    (\mathrm{all \, \, in \, \, limit} \, \, \, r \rightarrow 0) \label{B4} \nonumber
\end{align}
which gives
\begin{align}
&\alpha r H_{1}^{(1)}(\alpha r) J_{0}(k r) \sim-\frac{2 i}{\pi}   \\
 & k r H_{0}^{(1)}(\alpha r) J_{1}(k r) \sim \frac{i}{\pi}(k r)^{2} \ln (\alpha r) \sim 0 \label{B5}
\end{align}
and the Fourier transform as
\begin{equation}
\widetilde{H}_{0}^{(1)}(k, \alpha)=-\frac{4 i}{k^{2}-\alpha^{2}} \label{B6}
\end{equation}

If $k$ is extended from the positive real axis to the complex plane, $\widetilde{H}_{0}^{(1)}(k, \alpha)$ given by Eq. \eqref{B6}  has two poles $k_{\text {pole }}= \pm \alpha$. The two poles approach the real axis in the weak-damping $\left(\alpha^{\prime \prime} \downarrow 0\right)$ limit. The denominator in Eq. \eqref{B6}, in this limit, is approximately
\begin{align}
& k^{2}-\alpha^{2}=k^{2}-\alpha^{\prime 2}+\alpha^{\prime \prime 2}-2 \alpha^{\prime} \alpha^{\prime \prime} i \\
&\approx k^{2}-\alpha^{\prime 2}-2 \alpha^{\prime} \alpha^{\prime \prime} i \\
& =\left(k+\alpha^{\prime}\right)\left[k-\alpha^{\prime}-\frac{2 \alpha^{\prime} \alpha^{\prime \prime}}{k+\alpha^{\prime}} i\right]
\end{align}
and $\widetilde{H}_{0}^{(1)}(k, \alpha)$ tends, for real-valued $k>0$ and $\alpha^{\prime}>0$ (for out-going waves), to
\begin{align}
\lim _{\alpha^{\prime \prime} \downarrow 0} \widetilde{H}_{0}^{(1)}(k, \alpha) & =-\frac{4 i}{k+\alpha^{\prime}}\left[\wp \frac{1}{k-\alpha^{\prime}}+i \pi \delta\left(k-\alpha^{\prime}\right)\right] \\
& =2 \pi\left[\frac{1}{k} \delta\left(k-\alpha^{\prime}\right)-\frac{2}{\pi} \wp \frac{1}{k^{2}-\alpha^{\prime 2}} i\right] \label{B7}
\end{align}

Since $H_{0}^{(1)}\left(\alpha^{\prime} r\right)=J_{0}\left(\alpha^{\prime} r\right)+i Y_{0}\left(\alpha^{\prime} r\right)$, comparison with Eq. \eqref{B7} implies that the first term in the square bracket in Eq. \eqref{B7} is equal to the integral $\int_{0}^{\infty} d r r J_{0}(k r) J_{0}\left(\alpha^{\prime} r\right)$ and the second term $=i \int_{0}^{\infty} d r r J_{0}(k r) Y_{0}\left(\alpha^{\prime} r\right)$.


\section{Relation between skyrmion number and Berry phase}

\label{Sec:appendix-relation-chern}

The momentum space skyrmion number $Q_1$ discussed in Sec. \ref{Sec:Scalar-field-based-momentum-space-skyrmions} is different from the momentum-space linking number (MoLiN) in Ref. \cite{neuhaus-etal-1.2026arxiv} (under the assumption of so-called `spin-momentum locked waves' made in that reference), and therefore in this appendix we want to clarify the difference in more detail. First, as mentioned in the main text, the two quantities are clearly different, because in the case of a cylindrical skyrmion field the MoLiN is zero, while $Q_1$ is not necessarily zero.

In Ref. \cite{neuhaus-etal-1.2026arxiv} the MoLiN is said to be a Berry phase, while our $Q_1$ is a conventional skyrmion number (albeit in momentum space). It is therefore helpful to review the question how a skyrmion number can be related to concepts like Berry curvature, Berry connection, and - in the end - a Berry phase.

A quantum-mechanics approach to the relation between skyrmion and Berry quantities can be based on the interpretation of skyrmion vector fields as  Bloch vectors associated with the corresponding
Bloch Hamiltonians. Then, in the context of quantum mechanics, the well-known concept of Berry connections naturally provides a path to a Berry phase.

To proceed, let us write the skyrmion number introduced in Eqs. \eqref{I.1.a}-\eqref{I.1} and  \eqref{II.6}-\eqref{II.7} more generically as
\begin{equation}
Q^{skrym}=\frac{1}{4 \pi} \iint_{D} \mathrm{~d}^{2} k   \, \, \mathbf{n} \cdot\left(\frac{\partial \mathbf{n}}{\partial k_{x}} \times \frac{\partial \mathbf{n}}{\partial k_{y}}\right)
\label{eq:Qskrym-integral-01-appendix}
\end{equation}
where $D$ is the domain (here, for concreteness, taken to be in the two-dimensional momentum space), 
$\mathbf{n}(\mathbf{k})$ are real-valued unit vectors
and $\mathbf{k}=\left(k_{x}, k_{y}\right)$ is the wave vector in two dimensions.

The basic approach in the following will be to (1) interpret the vector field $\mathbf{n}(\mathbf{k})$  as a (normalized) Hamiltonian vector in a generic two-band Hamiltonian (analogous, for example, to the Dirac Hamiltonian of graphene), (2) use that Hamiltonian to define the Berry connection (via the 2-component complex Bloch functions), and (3) use the Berry connection to find  the Berry curvature.  
Integrating the Berry curvature over a compactified, boundary-less base space gives the first Chern number. For an integral of the Berry curvature over a simply-connected subset of the base space, if a gauge exists such that the Berry connection is smooth inside the subset, the integral gives the closed-loop Berry phase along the subset's boundary.

It is well known, for example in the context of topological insulators \cite{hasan-kane.2010,fruchart-carpentier.2013,shankar.2018arxiv}, that in Hermitian 2-band systems there is a relationship between the Chern number formulated in terms of the Berry curvature, which in turn is given by curl of the Berry connection, and the Chern number expressed solely in terms of the 2-band Hamiltonian 
\begin{equation}
H=\mathbf{h}(\mathbf{k}) \cdot \boldsymbol{\sigma}     \label{eq:Heqhdotsigma}
\end{equation}
(also often written as $\mathbf{d}(\mathbf{k}) \cdot \boldsymbol{\sigma}$)
where $\boldsymbol{\sigma}$ are the Pauli spin matrices. 
The relationship (equivalence) between the skyrmion number and the Chern number in the case of $2 \times 2$ matrices is well known, see for example Refs.  
 \cite{hasan-kane.2010,fruchart-carpentier.2013,cook.2023}, and also textbook discussions in 
 \cite{nakahara.2003,asboth-etal.2016,vanderbilt.2018}.
  The first Chern number is an integral over  the Berry curvature
\begin{equation}
    c_{1}=\frac{1}{2 \pi} \int_{D} d^{2} k\left(\frac{\partial A_{y}}{\partial k_{x}}-\frac{\partial A_{x}}{\partial k_{y}}\right)
    \label{eq:Chern-appendix}, 
\end{equation}
where the $A_{j}$ are the cartesian components ($j=x,y$) of the Berry connection for a given band $\nu$
\begin{equation}
\begin{aligned}
A_{j} & =i\left\langle u_{\nu, \mathbf{k}}\right| \frac{\partial}{\partial k_{j}}\left|u_{\nu, \mathbf{k}}\right\rangle 
\end{aligned}
\end{equation}
where $\nu$ is the band index  and  $\left|u_{\nu, \mathbf{k}}\right\rangle $ is the eigenfunction of Hamiltonian 
\eqref{eq:Heqhdotsigma}.
In Eq. \eqref{eq:Chern-appendix},
 we assume to deal with one of the two bands defined by the eigenenergies of $H$ (we omit the band index for simplicity), and $D$ is domain for $\left(k_{x}, k_{y}\right)$, which in the context of topological insulators is usually the Brillouin torus, but which in the present context 
is the compactified wave vector plane.
If the plane cannot be compactified, the integral in Eq. \eqref{eq:Chern-appendix} is not the Chern number (not a global topological invariant). 
Below, we will refer to the quantity $c_{1}$ as Chern-like integral, implying that it coincides with the Chern number only if the conditions for it to represent a global topological invariant are fulfilled.

Identifying the normalized Hamiltonian $\hat{\mathbf{h}}=\mathbf{h} /|\mathbf{h}|$ with the skyrmion vector field $\mathbf{n}$,
\begin{equation}
\mathbf{n}(\mathbf{k}) \equiv \hat{\mathbf{h}}(\mathbf{k})  \label{eq:n-eq-h}
\end{equation}
yields the desired relation between the skyrmion field and the Berry connection. The latter can be used to define a closed-loop Berry phase, 
\begin{equation}
    \gamma_{\cal{L}} = \oint_{\cal{L}} d \mathbf{k}  \cdot \mathbf{A}(\mathbf{k}) 
    \label{Eq:closed-loop-Berry-phase}
\end{equation}
where $\cal{L}$ denotes a closed loop (here in momentum space). Straightforward algebraic manipulation shows that the integrands of Eq. \eqref{eq:Qskrym-integral-01-appendix} and 
\eqref{eq:Chern-appendix} are equal, and therefore
\begin{equation}
c_1 = Q^{skyrm}
\end{equation}
regardless of the domain $D$. 
Owing to the gauge freedom of the Berry connection, after dividing by $2 \pi$, the closed-loop Berry phase, Eq. \eqref{Eq:closed-loop-Berry-phase}, can be chosen to be the same as the  
Chern-like integral,
see Eq. (3.37) of Ref. \cite{vanderbilt.2018}.

The skyrmion number, written as an integral over the skyrmion density, is most naturally discussed in terms of a geometric interpretation. It calculates the solid angle (in units of that of the unit sphere) subtended by the map $\hat{\mathbf{h}}(\mathbf{k})$ as $\mathbf{k}$ sweeps the entire domain $D$.

In topology, the degree of a map is usually defined for two compact oriented manifolds of the same dimension. It counts the number of times the domain manifold wraps around the range manifold, 
and the degree of a non-surjective map is zero.
(A more detailed discussion in the context of optical skyrmions is given in 
\cite{zhang-etal.2026skyrmions}).
However, in this paper we do not necessarily restrict the domain to a compact manifold, such as the Brillouin torus. In particular, we allow the domain to be the entire wave vector plane $\mathbb{R}^{2}$ 
and do not require a uniform limit of the map at large (infinite) radial distance from the origin.
Furthermore, while each unit vector $\hat{\mathbf{h}}$ lives on the sphere $S^{2}$, the map $\hat{\mathbf{h}}(\mathbf{k})$ may not cover the entire sphere an integer number of times.

The main step  used above is the usual construction of the Bloch vector (for a given band) as a normalized real-valued 3-component vector 
$ \mathbf{n} (\mathbf{k} )  \equiv \left\langle u_{1, \mathbf{k}}\right| \boldsymbol{\sigma} \left|u_{1, \mathbf{k}}\right\rangle  $. In other words, the normalized real-valued 3-component skyrmion  vector, which (due to the normalization condition $n_1^2 + n_2^2 + n_3^2 =1$)
is defined by two real parameters, is related to a quantum-mechanical problem defined by two-dimensional complex wave functions, such as $\left|u_{1, \mathbf{k}}\right\rangle$. Writing the wave functions generically as a vector $\left( 
\begin{array}{l}
z_1 \\
z_2
\end{array}\right)$
with complex components $z_i$, and noting that (i) the phase is arbitrary (we can, for example, choose $z_1$ to be real), and (ii) the normalization requires $|z_1|^2 +|z_2|^2 =1$, we see that the number of real parameters   defining the wave functions (here 2) equals that of the skyrmion vector. 
As a result of this construction (map), the skyrmion density is equal to the Berry curvature, and 
hence, after integrating the densities, the skyrmion number is equal to the 
Chern number.
 Formally, the underlying reason for the equality between the Chern and skyrmion number for the case of a $2 \times 2$ Hamiltonian is equivalence between the  complex projective space $CP^1$ and the sphere $S^2$. 
 For a discussion and generalization to higher dimensions, see Ref. \cite{cook.2023,liu-etal.2023}.

Coming back to the closed-loop Berry phase (MoLiN) in Ref. \cite{neuhaus-etal-1.2026arxiv},
we note that our analysis applies to the case of Hankel functions (including decaying Hankel functions with nonzero $\alpha''$ as well as the non-decaying limit $\alpha'' \downarrow 0$), whose imaginary part is not proportional to a delta-function and thus not a mathematical ring in momentum space, see Eqs. \eqref{II.1} and \eqref{II.2}. In other words, the momentum-space domain in  Ref. \cite{neuhaus-etal-1.2026arxiv}  is
reduced to a 1-dimensional ring $S^1$ with radius $k_0$ in momentum space, while ours is 2-dimensional.


\bibliographystyle{osajnl-no-comma}

\bibliography{skyrmion-number}



\end{document}